\documentclass[amsmat,amssymb,amsfonts,aps,prl,twocolumn,showpacs]{revtex4}

\usepackage{graphicx}
\usepackage{dcolumn}
\usepackage{bm}
\begin{document}

\title{Spin-transfer torque  on a single magnetic adatom   }
\author{F. Delgado,  J. J. Palacios, J. Fern\'andez-Rossier}
\affiliation{Departamento de F\'{\i}sica Aplicada,
Universidad de Alicante, San Vicente del Raspeig, 03690 Spain }

\date{\today}

\begin{abstract}

We theoretically show how the spin orientation of a single magnetic adatom can be controlled by spin polarized  electrons in a scanning tunneling microscope configuration.
The underlying 
physical mechanism  is spin assisted inelastic tunneling. By changing the direction of the applied 
current, the orientation of the magnetic adatom can be completely reversed on a 
time scale that ranges from a few nanoseconds to microseconds, depending on bias and temperature.
The changes in the adatom magnetization direction are, in turn, reflected in the tunneling conductance.

\end{abstract}

 \maketitle

There is now a fast growing interest  in controlling the spin orientation
of a single or few magnetic atoms in a solid state environment
for future spintronics and  quantum information devices.
So far this has only been achieved  using optical methods ~\cite{Le-Gall09}.
When a current flows through a magnetic region 
it becomes spin polarized due to exchange coupling between 
   transport electrons and localized magnetic moments.
The back action of transport electrons on the magnetic moment, known as spin-transfer torque~\cite{Slon}, 
can be used to rotate the magnetization in nanopillars made of billions of atoms~\cite{SMT-exp}. 
   The magnetization of such a large number of atoms  can be
 described with a single  classical vector and the current driven magnetization 
switching is properly  
modeled  by Landau-Lifsthitz equations 
extended  with the spin transfer term proposed by Slonczewski\cite{Slon}.   
Current induced magnetization switching  has been reported in much smaller nanomagnets, made of
100 atoms~\cite{SMT-nano}, but still in the semiclassical domain.

\begin{figure}
[hbt]
\includegraphics[angle=0,width=1\linewidth]{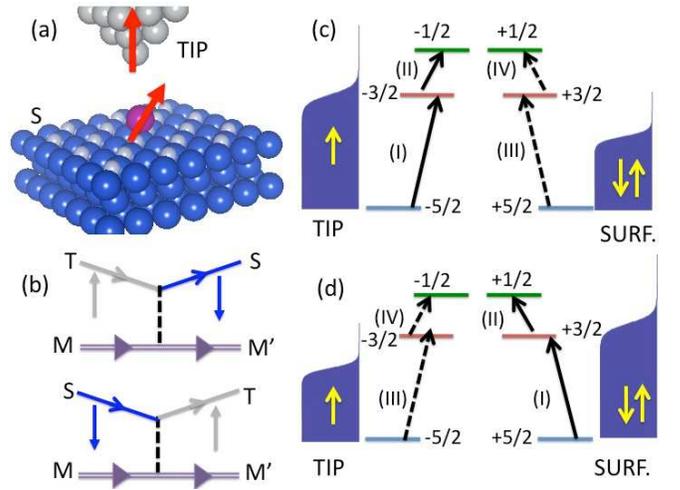}
\caption{ \label{fig1}(Color online). (a) 
Scheme of the proposed setup: a magnetic 
STM tip and a magnetic adatom on a insulating monolayer  deposited on a metal.
 (b) Two of the spin assisted tunneling events of Eq. (\ref{HTUN}). 
  Spin transitions when electrons flow from tip to surface ((c) and upper diagram in (b)) and
  from surface to tip ((d) and lower diagram in (b)). Dashed transitions  (III) and (IV)
are less efficient when the tip is polarized (see text).}
\end{figure}

       In the present letter we theoretically
show how a spin polarized current can be used to manipulate  the spin state of a single transition metal atom
deposited on a insulating monolayer on top of a metallic surface [see Fig.~\ref{fig1}(a)]. 
      The proposed implementation combines 
two alternative strategies used so far to probe the spin of a single atom on a surface:
     inelastic electron tunnel spectroscopy (IETS) with non-magnetic tips on one side and a spin polarized tunneling
current on the other.   The first technique
 has demonstrated that conveniently isolated  Mn, Fe and Co  adatoms
     have   quantized spin angular momentum along a well defined magnetic 
easy axis~\cite{Heinrich07,Heinrich08,Xue08,FePc}.  These experiments also  demonstrate that
      transport electrons and local spin are exchanged coupled \cite{JFR09, Fransson09, lorente}.  
In  contrast, experiments with ferromagnetic tips are based upon the sensitivity of conductance to the relative 
spin orientation of the tip and adatom (magnetoresistance)~\cite{Wiesendanger}.

Here we present a fully quantum mechanical theoretical analysis
 showing that, under the influence of spin polarized tunneling current, the spin of  a single magnetic atom can be directed either parallel or antiparallel to the  magnetic moment of the tip (or the surface).
When the current is  spin polarized along a given direction 
$\vec{n}$,  because either the tip or the substrate are ferromagnetic, a fraction of the tunneling electrons exchange one unit of spin with the magnetic adatom. This induces a net
transfer of spin along $\vec{n}$ towards the magnetic atom.
This current driven spin torque competes with the adatom spin relaxation provided by its weak coupling to the tip and substrate. 
The sign of the transfer is determined by the direction of the current whereas the efficiency is determined by the magnitude of the current.  

In order to study the spin dynamics of the magnetic adatom under the influence of the spin polarized tunneling electrons we derive and solve the master equation for the eigenstates $|M\rangle$ of the single ion spin Hamiltonian
\begin{equation}
{\cal H}_{\rm S}= D S_z^{'2} + E(S_x^{'2}-S_y^{'2}).
\end{equation}

Thus, the local spin is described quantum mechanically, 
in contrast to previous works~\cite{Nunez08,Fransson}. This  is crucial to account for the IETS experiments \cite{Heinrich07}. Here the prime denotes that the spin quantization axis is chosen with $z$ perpendicular to the surface. 
For convenience, we choose the quantization axis of the all the spins in the Hamiltonian  along $\vec{n}$ which makes it necessary to rotate ${\cal H}_{\rm S}$ when $\vec{n}$ is not perpendicular to the surface.
The value of the total spin $S^2$, and the magnetic anisotropy coefficients $D$ and $E$ change from atom to atom and also depend on the substrate \cite{Heinrich07,Heinrich08}. 

The scattering rates between states with different spin projection $|M\rangle$ 
arise from the exchange interaction of the adatom spin to delocalized electrons in tip and surface, 
hereafter denoted as the reservoirs and labeled with the indeces $\eta=T,S$. 
The total Hamiltonian of the adatom coupled to the reservoirs reads 
${\cal H}= {\cal H}_{\rm tip} + {\cal H}_{\rm sur} + {\cal H}_{\rm S}+ {\cal V} \;.$ 
The first three  terms describe the tip, surface and adatom as decoupled systems, whereas the ${\cal V}$ term
introduces interactions between them. 
The first term 
 describes the electrons in the ferromagnetic  tip magnetized along the direction $\vec{n}$, 
a static parameter in our theory. 
The second term is the Hamiltonian of the non-magnetic electrons of the metallic surface.
Thus, the quantum numbers of the transport electrons are  their momentum $k$,
spin along the $\vec{n}$ axis,  $\sigma$, and reservoir $\eta={\rm tip},{\rm sur}$.  
With this notation  we write
${\cal H}_{\rm tip} + {\cal H}_{\rm sur}=\sum_{k,\sigma,\eta} \epsilon_{\sigma\eta}(k)  
c^{\dagger}_{k\sigma\eta}c_{k\sigma\eta}\;$.
Since we consider a non-magnetic surface, we have 
$\epsilon_{\sigma,S}(k)=\epsilon_S(k) $.
  All the results of this paper are trivially  generalized to the case of a non-magnetic tip and a magnetic surface.

 The coupling of the atomic spin $\vec{S}$ and the conducting reservoirs has the 
form~\cite{Applebaum}
\begin{equation}
{\cal V}=
\sum_{\alpha,k,k',\sigma,\sigma',\eta,\eta'} 
T_{\eta,\eta',\alpha}(k,k')
\frac{\tau^{(\alpha)}_{\sigma\sigma'}}{2}  \hat{S}_{\alpha}
c^{\dagger}_{k\sigma\eta} c_{k'\sigma'\eta'}, 
\label{HTUN}
\end{equation}
where the index $\alpha$ runs over 4 values, $a=x,y,z$,   and  $\alpha=0$.
We use $\tau^{(a)}$ and $\hat{S}_{a}$ for the   Pauli matrices
and  the spin operators in the $\vec{n}$ frame, while $\hat{S}_{0}=I$ is the identity matrix.
$T_{\eta,\eta',\alpha}(k,k')$ for $\alpha=x,y,z$ is the exchange-tunneling interaction between the localized
spin and the transport electrons and potential scattering for $\alpha=0$.
Attending to the nature of the initial and final electrode, 
Eq. (\ref{HTUN}) describes four types of exchange interaction, two of which contribute to the 
current, the other two conserving the charge difference between tip and surface. 
The former are crucial to account for the magnetic IETS~\cite{JFR09} and tend to ``heat'' the spin of the adatom. The other two  provide an efficient cooling mechanism, through a Korringa like spin relaxation, and were not included in previous work \cite{JFR09,Fransson09,lorente}. 
Following Anderson ~\cite{Anderson66},  we assume that (\ref{HTUN}) arises from kinetic exchange.
The momentum dependence of $T_{\eta,\eta',\alpha}(k,k')$
can have important consequences in  the conductance
 profile~\cite{Merino_Gunnarsson_prb09} in a energy scale of $eV$, so it can be safely neglected in IETS. 
We thus parametrize $T_{\eta,\eta',\alpha=0}(k,k')= v_{\eta}v_{\eta'}  T_{0}$ and
$T_{\eta,\eta',a}(k,k')= v_{\eta}v_{\eta'}  T_S$, where $ T_S$ is the same for $a=x,y,z$ and $v_{\rm sur}$ 
and $v_{\rm tip}$ are  dimensionless factors that scale as
 the surface-adatom and tip-adatom
hopping integrals.  This parametrization sets a relation between spin torque and spin relaxation.

The  occupation of the spin states $|M\rangle$,  $P_M$,  are governed by
 the master equation
\begin{equation}
\frac{dP_M}{dt}=\sum_{M',\eta\eta'}P_{M'}W_{M',M}^{\eta'\to \eta}-P_M\sum_{M',\eta\eta'}
W_{M,M'}^{\eta\to \eta'},
\label{master}
\end{equation}
where 
$W_{M,M'}^{\eta\to\eta'}$ are the scattering rates from state $M$ to $M'$  induced by interaction with a quasiparticle which is initially in reservoir $\eta$ and ends up in $\eta'$.  Eq. (\ref{master}) does not include spin coherences. This approximation is good provided that spin decoherence is faster than spin relaxation, which is known to be the case due to hyperfine coupling \cite{Le-Gall09} in Mn atom.  
After some algebra, the scattering rates can be written as:
\begin{eqnarray}
W_{M,M'}^{\eta\to \eta'}
= \frac{\pi |T_S v_\eta v_{\eta'}|^2}{\hbar}
{\cal G}(\Delta_{M,M}
+\mu_\eta-\mu_{\eta'})
\Sigma_{M,M'}^{\eta\eta'},
\label{ratenew}
\end{eqnarray}
where   ${\cal G}(\omega)\equiv
\omega\left(1-e^{-\beta \omega}\right)^{-1} $ are the phase space factors associated to quasiparticle scattering,
$\Delta_{M,M'}=E_M-E_{M'}$ is the energy change of the atom, $\mu_\eta$ is the chemical potential of electrode $\eta$ and 
$\Sigma_{M,M'}^{\eta,\eta'}$ are spin matrix elements 
\begin{eqnarray}
&&2\Sigma_{M,M'}^{\eta\eta'}=\left|  {\cal S}_z^{M,M'}  \right|^2 
(\rho_{\eta\uparrow} \rho_{\eta'\uparrow}
+\rho_{\eta\downarrow} \rho_{\eta'\downarrow})
\crcr
&&\;\;+\left|{\cal S}_+^{M,M'} \right|^2 \rho_{\eta\downarrow}\rho_{\eta'\uparrow} +
\left|{\cal S}_-^{M,M'}\right|^2\rho_{\eta\uparrow}\rho_{\eta'\downarrow}
\label{sigma}
\end{eqnarray}
where ${\cal S}_a^{M,M'}= \langle M|S_a|M'\rangle$
and $\rho_{\eta,\sigma}$ is the density of states at the Fermi energy for spin $\sigma$ in the electrode $\eta$.
These equations show that,  for a ferromagnetic tip, spin flip and spin conserving  rates  are different, in contrast to the case of a non-magnetic tip~\cite{JFR09}. 
The rates that induce changes in the adatom spin population are all proportional to $T_S^2$, and can  be classified in three groups: intra-tip,  intra-surface  and tip-surface rates ($\propto v_{\rm tip}^4$ , 
$\propto v_{\rm sur}^4$, and $\propto v_{\rm tip}^2 v_{\rm sur}^2$, respectively).  In the first two, 
the capability to transfer energy to the magnetic atom is given by the temperature, being the release of energy from the atom to the electrodes always allowed.  In contrast, the rates carrying current  
can transfer energy to the atom even at zero temperature, provided that there is a bias voltage $eV=\mu_S - \mu_T$.   

On top of the $ T_S^2$ rates, there are scattering processes of order $ T_0^2$ and 
$T_0T_S$ that do not change the occupations 
but contribute to  the current.  Evaluated to the same second order in ${\cal V}$ than the rates in the master equation, the expression of the current has three terms,   $I=I_0 + I_{MR} + I_{IN}$.  The first term is elastic  and spin independent.  
The second is the elastic  but
 sensitive to the relative spin orientation of the adatom and the tip magnetic moment. 
The third term comes from the inelastic exchange processes being proportional to $T_S^2$.
The expressions for the three contributions to the current, analogous to those derived elsewhere \cite{Fransson09}, are:
\begin{eqnarray}
&&I_0+I_{MR}=-\frac{2}{e} G_0\left(1+ x \langle S_z\rangle {\cal P}_T\right)i_-(-eV))
\label{current1}
\\
&&I_{IN}=-\frac{G_{S}}{e} \sum_{M,M'}
\Big[ i_{-}(\Delta_{M,M'}-eV) \sum_a \left| {\cal S}_a^{M,M'}\right|^2
 \nonumber \\
&&
+ {\cal P}_T i_{+}(\Delta_{M,M'}-eV)
{\rm Im}\left({\cal S}_x^{M,M'}{\cal S}_y^{M',M} \right)\Big]P_M(V)\;\;
\label{current2}
\end{eqnarray}
Here $G_0\equiv \frac{e^2\pi\rho_T\rho_S}{4\hbar}\left| T_0 v_{\rm tip} v_{\rm sur}\right|^2$
is  the elastic conductance of the junction while  $G_S=x^2 G_0$, with $x=T_S/T_0$
the  relative intensity of the inelastic channel.
We define the spin polarization of the tip ${\cal P}_T= \left(\rho_{T\uparrow}-\rho_{T\downarrow}\right)/\left(\rho_{T\uparrow}+\rho_{T\downarrow}\right)$ . Functions $i_\pm$ are defined as
 $i_{\pm}(\Delta_{M,M'}-eV)= {\cal G}(\Delta_{M,M'}-eV) \pm {\cal G}(\Delta_{M,M'}+eV)$.
The average adatom magnetization along the tip magnetization axis is    $\langle S_z \rangle=\sum_M P_M(V) \langle M|S_z|M\rangle $ and depends on the bias voltage, the central result of the manuscript.
Importantly,  the magnetoresistive contribution to the elastic current  permits to track changes in magnetization. 
Although $x$ probably varies from system to system, 
 we take $x=1$, close to the results reported for a single Mn adatom \cite{Heinrich07}

The inelastic contribution has also two terms, one of which depends on the tip polarization. 
The first  inelastic term is proportional to $i_{-}$, whose derivative with respect to $V$  
  gives the characteristic  inelastic conductance   steps  
as  $|eV|$  goes across  $\pm \Delta_{M,M'}$.  By contrast, the ${\cal P}_T$ dependent term of the inelastic conductance, proportional to  $i_+$, yields  steps at the excitation energies of opposite sign as the polarity of the bias is reversed.  Both the elastic and inelastic term proportional to ${\cal P}_T$ can produce a $dI/dV$ which is not an even function of bias. 

\begin{figure}
[t]
\includegraphics[width=0.9\linewidth,angle=0]{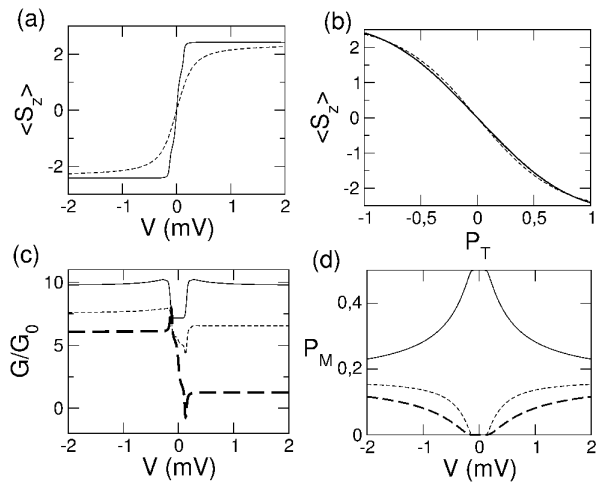}
\caption{ \label{fig2}(a) Average magnetization $\langle S_z\rangle$ versus applied bias for $T=1$K
(dashed line) and $T=0.1$K (solid line). (b) Average magnetization after saturation ($eV=5meV$)
versus tip polarization for the same two temperatures.
 (c) $dI/dV$ for single Mn in Cu$_2$N surface probed with a non-polarized tip (solid line) and two
different tip polarizations, ${\cal P}_T=-1/3$ (thin-dashed line) and ${\cal P}_T=-1$ (thick-dashed line) for a fixed temperature $T=0.1$K. (d) Variation of the populations $P_M$ with applied bias for non-polarized tip and $T=0.1$K.}
\end{figure}

We now can address the main question of this manuscript: How does a spin polarized tunneling current  affect the magnetization of a single adatom?.   We consider first the simplest possible situation, where the magnetic moment of the tip is parallel to the easy axis of the single atom. 
We
choose a single Mn atom on a Cu$_2$N surface, which in the case of non-magnetic tip is very well characterized experimentally \cite{Heinrich07} 
and theoretically\cite{JFR09,Fransson09,lorente,Shick_Maca_prb2009}.  The spin of the Mn atom 
in this situation is $S=5/2$ with $D=-0.039$ meV and $E=0.007$ meV.  Since $E<<|D|$
we can limit our qualitative discussion to the case $E=0$, so that  the eigenstates of ${\cal H}_S$ are also eigenstates of $S_z$. 
 The numerical simulations are done with $E\not =0$  and do not change qualitatively.
  In the absence of applied magnetic field and at temperatures much smaller than the zero field splitting $4|D|$, the equilibrium distribution is such that the two ground states, $S_z=\pm5/2$,  are equally likely and the equilibrium average magnetization is zero.  

The non-equilibrium dynamics of the atom depends on a number of parameters that can be tuned experimentally, such as the bias voltage,  the temperature, and the ratio $r\equiv\frac{v_{\rm tip}}{v_{\rm sur}}$.    The latter depends on the tip-adatom distance.  As a general rule,  the processes that drive the magnetic adatom out of equilibrium are proportional to $v_{\rm tip}^2v_{\rm sur}^2$ whereas the processes that cool the spin down (provided that $k_bT <e|V|$) are proportional to $v_{\rm tip}^4+v_{\rm sur}^4$.  
 Thus, the non-equilibrium effects are higher as $r$ increases. We always take $r<1$.
 Other parameters, such as
 the inelastic ratio $x$ and the magnitude of the tip polarization ${\cal P}_T$ are not so easy to control.

In Fig. \ref{fig2}(a) we plot the {\em steady state} average magnetization of the Mn atom
for ${\cal P}_T=-1$ and $v_s=2v_T=1$ and two temperatures, $T=1$K and $T=0.1$K, as obtained from solving Eq. (\ref{master}). The
tip polarization was assumed parallel to the Mn easy axis.
The result has three outstanding features. First, the magnetization of the atom can be reversed from parallel to antiparallel just by electrical
 means [Fig. \ref{fig2}(a)]. Second, the saturation magnetization can take very large values, which increase up to 100 percent for $E=0$ as the degree of spin polarization of the tip increases [see Fig.\ref{fig2}(b)]. Third, the magnetization switch produces a large assimetry between positive and negative bias   conductance
  $G=dI/dV$ [Fig. \ref{fig2}(c)] which would be the experimental smoking gun of the spin transfer.

The  magnetization shown in Fig. 2(a,b)  arises from the competition between spin-transfer from the spin polarized current to the atom and  spin relaxation. The microscopic
spin-transfer events, depicted in Fig. 1(b,c,d) , are the same than those resulting in steps in IETS \cite{Heinrich07,JFR09}. In the case of Mn atom probed by non-magnetic tip,  a single step has been reported \cite{Heinrich07} , related to  the spin increasing transitions $-5/2 \rightarrow -3/2$ and its time reversal counterpart, the  spin decreasing  transition $+5/2\rightarrow +3/2$, both  with energy $4|D|$  (see Fig. (\ref{fig1}c,d)).  For non magnetic electrodes there is no net spin transfer because  the rates (\ref{ratenew}) of these two processes are identical. In contrast,  net spin transfer occurs when the tip is magnetic and these two processes have different rates.    This is more easily seen
in a  extreme case  ${\cal P}_T=1$,  a half metallic tip such that 
$\rho_{T\downarrow}=0$.  Then, spin flip assisted tunneling from tip to surface can only  decrease the transport spin (upper diagram in Fig. 1b), {\em increasing} the spin of the magnetic adatom  (process (I) and (II)  in Fig. (1c)) whereas processes (III) and (IV) (Fig. 1c) are not allowed, or for non-fully polarized tips, partly suppressed.  When the bias is reversed, electrons flow from the  surface to the tip.  In this case, $\rho_{T\downarrow}=0$ implies that surface-tip 
  spin-flip assisted tunneling can only increase the transport spin (lower diagram in Fig. 1b), {\em decreasing} the adatom spin (processes (I) and (II) in Fig. 1d).


 In the case of non-magnetic tip, our theory predicts a non-equilibrium effect which has been already observed experimentally \cite{Heinrich08}: the small decay of $dI/dV$  for $eV$ larger than the inelastic threshold. 
   The spin-flip assisted events (Fig. 1(b,c,d))  deplete the ground state doublet  in favor of the first excited state doublet [fig. (\ref{fig2}d)].  This opens two new transport channels ($\pm 3/2 \rightarrow \pm 1/2$, (processes (II) and (IV) in Fig. 1(c,d)) which, in this case, happens to be  less efficient than the channels
$\pm 5/2 \rightarrow \pm 3/2$ (processes (I) and (III) in Fig. 1(c,d)). As the bias increases above $4|D|$,  the efficient inelastic transitions (I,III) are replaced by the less efficient transitions (II,IV),   decreasing of the inelastic contribution to  $dI/dV$.

\begin{figure}
[hbt]
\includegraphics[height=0.8\linewidth,width=0.6\linewidth,angle=-90]{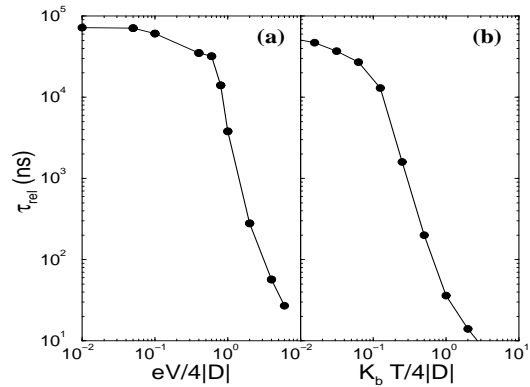}
\caption{ \label{fig3}Switching time $\tau_{rel}$ for ${\cal P}_T=-1$, $v_S=1$ and $v_T=0.5$.
(a) $\tau_{rel}$ as a function of applied bias $V$ for $T=0.1$K and (b),
 as a function of temperature for $eV=2|D|$.}
\end{figure}

The spin transfer events that "heat" the adatom, require that  a quasiparticle has an excess energy
 larger than  $4|D|$. They compete with spin cooling events (order $v_{\rm tip}^4, v_{\rm sur}^4$) that are  always available. 
At $T=0$,  the current induces changes in the magnetization only for  bias larger  than the inelastic threshold $4|D|$. At finite temperature there are always 
  quasiparticles thermally activated  above the $4|D|$ 
threshold and they are enough to induce the magnetization in the atom, even for $eV<4|D|$,  provided that a sufficiently long time is used in the process. In Fig.~\ref{fig3} we show how the time-scale necessary for the switching   depends dramatically both on $eV$ and $k_bT$: the switching time can decrease by up to four orders of magnitude as either the bias goes above $4|D|$ or the temperature is raised above $4|D|/k_b$. 

Finally, we have studied the   effect of the deviations of the magnetization
axis $\vec{n}$ with respect to the adatom easy axis. Curves similar to Fig. 2a are obtained for the average magnetization of the adatom along $\vec{n}$ as a function of $eV$, but the saturation magnetization at high bias is a decreasing function of the angle formed by $\vec{n}$ and the adatom easy axis. Whereas a small tilt (5 degrees) barely changes the curve compared to Fig. 2a,   the efficiency
of the spin transfer is minimal when $\vec{n}$ is perpendicular to the adatom easy axis but still can reach 0.6.

  In conclusion we have shown that the spin of a single magnetic adatom can be  polarized,   reversed and monitored by means of spin polarized currents in the STM configuration.  The time scale for the single spin switching can be as quick as a few nanoseconds.  
     Our proposal adds the possibility of  single spin control to the wide range of uses of STM in the field of  nanospintronics~\cite{Heinrich07,Heinrich08}. 
This possibility is based on non-equilibrium processes and differs from recent work in which switching is achieved by a change of sign of the equilibrium  tip-adatom exchange interaction \cite{Bruno09}.

We acknowledge fruitful discussions with A. Levy-Yeyati, C. Cuevas, C. Untiedt and  S. De Francheschi.  This work has been financially supported by MEC-Spain (Grants  MAT07-67845  and  
CONSOLIDER CSD2007-00010.  FD acknowledges the Spanish Program "Juan de la Cierva".



\newpage

\end{document}